\documentstyle[preprint,aps,epsfig]{revtex}
\begin{document}
\draft
\title{Cross Section Fluctuations of  Photon Projectile
in Generalized Vector Meson Dominance Model }
\author{L. Frankfurt}
\address{School of Physics and Astronomy,\\
Raymond and Beverly Sackler Faculty of Exact Sciences,\\
Tel Aviv University, 69978 Tel Aviv, Israel\\
and Inst. of Nuclear Physics, St. Petersburg, Russia}
\author{V. Guzey}
\address{Department of Physics\\
The Pennsylvania State University, University Park, PA 16802}
\author{and\\
M. Strikman}
\address{Department of Physics\\
The Pennsylvania State University, University Park, PA 16802\\
and Inst. of Nuclear Physics, St. Petersburg, Russia}
\maketitle
\begin{abstract}

We explain that the generalized vector meson dominance (GVMD) model is 
successful in describing hard high energy processes  because it
contains  expected in QCD Color Coherent Phenomena. Within the model,
which reproduces Bjorken scaling and nuclear 
shadowing for deep inelastic processes, we calculate 
$P_{\gamma}(\sigma,Q^2)$, the probability for the photon to interact with 
a target with the cross section $\sigma$. We find that 
within the GVMD model the virtual
photon has two main components: the ''hard'' one with the small cross 
section of the order of $1/Q^2$ and the high probability and the ''soft'' 
one with a typical hadronic cross section with the probability 
suppressed by the factor of $1/Q^2$. 
Thus the GVDM model produces  $P_{\gamma}(\sigma,Q^2)$ similar to that
within the parton model and suggests the conspiracy between 
hard and soft QCD at $Q^2 > 1$ GeV$^2$.
  
We  study the limit $\sigma \rightarrow 0$ and show, that 
$P_{\gamma}(\sigma,Q^2) \propto 1/ \sigma$ in both GVMD model and QCD.
Thus similarly to the parton model the GVMD model predicts the 
Color Transparency phenomenon. This is an indirect indication that 
Bjorken scaling, Color Transparency and shadowing in nuclei
in DIS are closely related.

\end{abstract}
\narrowtext
\section{Introduction}

The aim of this paper is to demonstrate that a vector 
dominance model, generalized to comply with Bjorken scaling, (GVMD),
contains Color Coherent Phenomena anticipated previously
within the parton model and/or for hard PQCD regime. 
 This explains the ability of GVDM
to describe large sets of data and demonstrates once more  that 
small $x$ physics of deep inelastic processes is a nontrivial interplay of
soft and hard QCD physics.

Simple vector meson dominance (VMD) models, for review see 
\cite{Fey}, \cite{Bauer}, \cite{Shaw-book} and references therein, 
describe successfully processes involving photon-hadron interactions 
at low ''photon mass'' $Q^2$. However such models, containing a 
finite number of vector mesons, do not reproduce approximate Bjorken 
scaling of $\sigma_{\gamma^{\ast} N}$ at large $Q^2$.

In order to restore the property of Bjorken scaling within GVDM, the 
photon was represented as an infinite sum of vector mesons 
\cite{Shaw-book}. The convergence of the sum within this model
requires that the cross section of the 
meson-nucleon interaction decreases with the mass of the meson. In 
such models nuclear shadowing dies away as $Q^2$ increases, which 
contradicts recent experiments \cite{exp1}, which reveal only a 
weak dependence of $F_{2A}(x,Q^2)$ on $Q^2$.

This problem does not occur in generalized vector dominance (GVMD)
models, where non-diagonal transitions among mesons have been
introduced. The model we use conjectures that there are transitions 
only between mesons with neighboring masses \cite{Schildknecht}, \cite{Shaw}.
The scattering matrix, {\bf S}, then, acquires non-diagonal 
elements in the basis of vector mesons. The model describes reasonably
well  Bjorken scaling of $\sigma_{\gamma^{\ast} N}$ 
\cite{Schildknecht}, \cite{Shaw} and  nuclear shadowing in DIS \cite{Shaw}.

 Alternatively, one can expand the photonic state through states with 
the particular  cross section of interaction with a target
\cite{Walker}. 
It is so called  method of cross section fluctuations. The advantage 
of such an approach is that different states, having different cross 
section of interaction, do not mix, therefore the scattering matrix
{\bf S} or {\bf T} is diagonal in such a basis. 

The physical ground for this approach is the fact that 
configurations of a  different geometrical size are 
present in the photon. It is known that the strength 
of the interaction of such a configuration depends 
on its geometrical size, consequently the photon can 
be represented as a superposition of eigenstates of 
the {\bf T}-matrix, $|\psi_{n}\rangle$:
\begin{equation}
|\gamma \rangle=\sum c_{n}|\psi_{n} \rangle \label{dec},
\end{equation}
which interact with the target with different strengths:
\begin{equation}
{\it {\bf T}}|\psi_{n} \rangle=\sigma_{n}|\psi_{n} \rangle .
\end{equation}
It is convenient to introduce the distribution over cross 
sections $P_{\gamma}(\sigma)$, which gives the probability 
for the photon to interact with the target with the cross 
section $\sigma$. Having found the set of eigenvalues $\sigma_{n}$
and corresponding coefficients $c_{n}$, one can reconstruct 
the distribution $P_{\gamma}(\sigma)$ according to the rule
\cite{Walker},\cite{rule}:
\begin{equation}
P_{\gamma}(\sigma)=\sum_{n} |c_{n}|^2 \delta (\sigma-\sigma_{n}) \label{P} .
\end{equation}

There are no methods allowing for a calculation
of the distribution $P_{\gamma}(\sigma,Q^2)$ from the first principles, 
except for $\sigma \rightarrow 0$. In the present paper we evaluate 
$P_{\gamma}(\sigma,Q^2)$ within the framework of the GVMD model
and find conspiracy between hard and soft physics
suggested initially within the parton model \cite{Bjorken}: 
although the effective cross section decreases as ${1/Q^2}$,
the probability of configurations interacting with a typical hadronic 
cross section is ${1/Q^2}$ also.
Thus the significant contribution of non-diagonal transitions resolves the
Gribov puzzle \cite{Gribov}--the  contradiction of pre-QCD ideas to 
 approximate Bjorken scaling for deep inelastic processes. 
 
We  study  
asymptotic properties of $P_{\gamma}(\sigma,Q^2)$ at small $\sigma$ and 
compare predictions obtained in the GVMD model and QCD.

Once again we want to point out that the very fact of existence  of 
the distribution over cross sections $P_{\gamma}(\sigma,Q^2)$ proves that 
there are configurations of different strengths in the photon. 
         
\section{Evaluation of $P_{\gamma}(\sigma,Q^2)$ within GVMD model }

At high energies a photon interacts with a target by means of its 
hadronic components. The generalized vector meson dominance model 
that we use \cite{Schildknecht}, \cite{Shaw} can be expressed as the following decomposition:
\begin{equation}
|\gamma\rangle=\sum_{n}^{\infty} \frac{e}{f_{n}} \frac{M_{n}^2}
{M_{n}^2+Q^2} |\rho_{n}\rangle \label{vmd}.
\end{equation}
Parameters of the model are chosen such that
\begin{equation}
M_{n}^2=M_{0}^2 (1+2n) ,
\end{equation}
where  $M_{0}$=0.77 GeV, assuming $|\rho_{0}\rangle=|\rho\rangle$, and 
\begin{equation}
\frac{M_{n}^2}{f_{n}^2}=\frac{M_{0}^2}{f_{0}^2} ,
\end{equation}
where $f_{0}/4\pi$=2.36 .

Note that 
for simplicity we consider only one kind of
vector mesons here, discarding contributions of $\omega$ and $\phi$ mesons.
The inclusion of additional flavors would not change qualitative conclusions
of this paper.

The total photoabsorption cross section for transverse photons on 
nucleons is \cite{Schildknecht}, \cite{Shaw}: 
\begin{equation}
\sigma_{T}(s,Q^2)=\sum_{nm} \frac{e}{f_{n}}
\frac{M_{n}^2}{M_{n}^2+Q^2}
\Sigma_{nm}\frac{e}{f_{m}} \frac{M_{m}^2}{M_{m}^2+Q^2} \label{ttot}.
\end{equation}
$\Sigma_{nm}$ is the scattering matrix in the basis of vector mesons. 
Assuming that there are transitions only between the mesons with 
neighboring masses, this matrix takes up a symmetrical 
tridiagonal form \cite{Schildknecht}, \cite{Shaw} :
\begin{equation}
\Sigma_{nm}=\sigma_{0}\delta_{nm}-\frac{\sigma_{0}}{2}
\frac{M_{n}}{M_{n+1}}(1-0.3\frac{M_{0}^2}{M_{n}^2})\delta_{nm\pm1} , 
\label{sigmaexpl}
\end{equation}
where $\sigma_{0}$=25 mb is the total cross section of interaction of 
vector mesons with the target. 

Non-diagonal elements of the matrix  $\Sigma_{nm}$ are chosen in such
a way that at large masses $M_{n}$  their first terms
cancel diagonal elements. Their second terms have the form of a 
simple (nongeneralized) 
vector meson dominance model with the cross section inversely 
proportional to the mass squared of the mesons. Thus, the model 
\cite{Schildknecht}, \cite{Shaw} reproduces approximate Bjorken scaling at relatively 
large $Q^2$ in spite of the fact that the cross section of the 
diagonal transitions does not decrease with $M^2_{n}$.

We want to stress that within  conventional approaches such as the 
parton model, PQCD, nonrelativistic quark models of a hadron,
cross sections of the scattering of excited hadronic states
off a hadron target are not very different (may be larger) than 
the cross section of the ground state off the same target.

The scattering matrix {\bf T} is diagonal in the basis of 
eigenvectors  $|\psi_{n}\rangle$ with eigenvalues $\sigma_{n}$ 
giving possible cross sections. Thus, the problem of relating 
the two formalisms reduces to finding
eigenvalues and eigenvectors
of $\Sigma_{nm}$. Having found the representation of the vector of 
state of each  meson in the basis of eigenvectors, and 
using \ (\ref{vmd}), one can find coefficients $c_{n}$ 
in decomposition \ (\ref{dec}) and, thus, reconstruct  
$P_{\gamma}(\sigma,Q^2)$ according to \ (\ref{P}).

The numerical solution to the problem was carried out 
with help of the computer software {\it Mathematica} for 10 mesons. 
The graphic solution is given in Fig. 1 for $Q^2$=0, 1, 2 GeV$^2$ (we plot $f_{0}^2/ e^{2} \cdot P(\sigma,Q^2$)). 
For such $Q^2$ it is sufficient to take into account only first ten mesons. 

  A more universal characteristic, which depends weakly on a 
number of vector mesons, is  moments of the 
distribution $P_{\gamma}(\sigma,Q^2)$ defined as:
\begin{equation}
\langle \sigma^{n} \rangle=\int P_{\gamma}(\sigma,Q^2) \sigma^{n} d \sigma .
\end{equation}
We give first five moments computed for our specific example of 
10 vector mesons in Table \ (\ref{Table1}), although the moments are 
not sensitive to the number of mesons taken into consideration.

{}From Fig. 1 one can see the following general tendencies of behavior 
of  $P_{\gamma}(\sigma,Q^2)$:
\begin{enumerate}
\item{$P_{\gamma}(\sigma,Q^2) \propto 1/ \sigma$ at small $\sigma$}
\item{$P_{\gamma}(\sigma,Q^2) \rightarrow 0$ at large $\sigma$}
\item{$P_{\gamma}(\sigma,Q^2)$ decreases {\it on average} with 
increase of  $\sigma$}
\item{$\langle \sigma^{n} \rangle \propto 1 / Q^2$ for n=1,2 \dots}
\end{enumerate}

Generally speaking, though the exact analytical form of the
distribution $P_{\gamma}(\sigma,Q^2)$ depends on a number of vector 
mesons taken into account ( on the dimension of $\Sigma_{nm}$), the 
general tendencies given above reflect universal properties of  the 
distribution $P_{\gamma}(\sigma,Q^2)$.

For calculation purposes it is useful to have an analytical
expression of $P_{\gamma}(\sigma,Q^2)$ defined for all $\sigma$. We
suggest the following parameterization, which reflects the general 
properties of $P_{\gamma}(\sigma,Q^2)$:
\begin{eqnarray}
P_{\gamma}(\sigma,Q^2)&=&N\Big(\frac{1}{\sigma/\sigma_{0}} \nonumber\
\Theta(\frac{1}{C} \frac{\mu_{1}^2}{\mu_{1}^2+Q^2}- \sigma / \sigma_{0} )+\frac{\mu^2}{\mu^2+Q^2}\Pi(\sigma/\sigma_{0}) \Big) \nonumber\\
\Pi(\sigma/\sigma_{0})&=&6.03 \, exp(-15.15 \cdot(\sigma/\sigma_{0}-0.7)^2) \nonumber\\
\mu_{1}^2&=&0.32  \ {\rm GeV}^2 \nonumber\\
\mu^2&=&0.39 \ {\rm GeV}^2 \nonumber\\
C&=&0.589 \nonumber\\
N&=&0.387 \label{param}
\end{eqnarray}
Here $\Theta$ is a step--function. The presented parameterization 
reproduces first three moments of the actual  distribution 
$P_{\gamma}(\sigma,Q^2)$ quite accurately for 0 $< Q^2 <$ 1  GeV$^2$:
the total cross section is given with the accuracy of 4\%, the 
second moment is fitted with the accuracy of 8\%, the third 
moment accuracy is at the level of 15\%.

One can see that this distribution contains a nontrivial interplay 
between hard and  soft physics. The first term is a ''hard'' piece 
with  the cross section proportional to $1/Q^2$ and the probability  
which does not depend on $Q^2$. The second term is a ''soft'' piece 
with the large cross section and the probability which dies away as 
$1/Q^2$. While both terms contribute to the total cross section, the 
dominant contribution to higher moments comes from the soft part of 
$P_{\gamma}(\sigma,Q^2)$ which guarantees that nuclear shadowing 
is the leading twist effect. One can also see it from  discrete 
versions of the distribution  $P_{\gamma}(\sigma,Q^2)$ for 10 and 20 
vector mesons.  

We have shown that this particular GVMD model leads to the existence
of the nontrivial distribution over cross section, which agrees with 
the notion that the photon consists  of $q\bar{q}$ configurations, 
having different geometrical sizes, therefore interacting with
different cross sections. This phenomenon is called 
{\it cross section fluctuations}.

\section{Nuclear shadowing}
In the total cross section of deep inelastic $\gamma^{\ast} \, A$ 
scattering nuclear shadowing is predominantly  inelastic shadowing, 
that is due to high mass intermediate states \cite{Gribov}. The 
leading contribution (double rescattering) can be expressed in terms 
of the cross section of inclusive forward diffractive dissociation 
of the virtual photon into states ''$X$'':
\mbox{$\gamma^{\ast} +N \rightarrow X+N$}.

The natural assumption, that any hadron state interacts with 
sufficiently heavy nuclei with the same cross section 
$\pi R_{A}^2$, leads to the Gribov relationship
between the total photoabsorption cross section and cross
section of the process $e \bar{e} \rightarrow hadrons$,
which contradicts  Bjorken scaling \cite{Gribov2}. 
The idea how to resolve the Gribov puzzle has been suggested 
by Bjorken \cite{Bjorken}.
He applied the parton model to the light cone wave function of the
energetic photon and concluded  
that the momentum of the quark (antiquark) in the photon should
be aligned along the momentum of the photon. Such rare,-- with the 
probability $1/Q^2$,--large size, asymmetric configurations 
give the dominant contribution into the cross section --the  aligned jet model 
\cite{Bjorken}. 
 Thus most of quark-antiquark  configurations in the photon
are sterile within the parton model approximation.

In QCD this picture is 
modified by identifying sterile states as 
colorless quark-gluon configurations of a spatially 
small size having small interaction cross 
sections and by including QCD evolution of sterile states.
QCD evolution leads to a fast increase with energy of the cross section 
of interaction of quark configurations having large $k_{t}$ and
to the parton bremsstrahlung \cite{physrep88}. 
Thus the QCD improved  aligned jet model predicts a nontrivial 
competition between 
two contributions into $P_{\gamma}(\sigma,Q^2)$. The  soft piece
corresponds to a usual hadronic cross section but with the 
probability $1/Q^2$. The hard piece to $P_{\gamma}(\sigma,Q^2)$ comes from 
cross sections $\sigma \propto 1/Q^2$, but their probability
does not decrease with $Q^2$ .

Another useful representation of  the 
cross section of inclusive forward diffractive dissociation 
of the virtual photon into states ''$X$'' and 
deep inelastic $\gamma^{\ast} \, A$ scattering can be given 
in terms of cross section fluctuations \cite{MP}.
Within this representation nuclear shadowing in DIS is a  
consequence of significant cross section fluctuations. 
Nuclear shadowing in DIS is given by the series of terms such 
as $\sigma^{n}$. The first term in this series is  
\mbox{$\propto \langle \sigma^{2} \rangle / \langle \sigma \rangle$} 
\cite{MP},\cite{KL}.

The diagonal vector meson dominance model assumes that cross section 
of the interaction of hadronic components of the photon falls off 
as the inverse mass squared of the component, $\sigma_{h}\propto 1/m^2_{h}$.
So within this model
$\langle \sigma \rangle \propto 1/Q^2$, 
$\langle \sigma^2 \rangle \propto ln(Q^2)/Q^4$, which makes
\begin{equation}
\frac{\langle \sigma^2 \rangle}{\langle \sigma \rangle}=\frac{ln(Q^2)}{Q^2}.
 \end{equation}
 
Therefore within this model nuclear shadowing dies out at large $Q^2$ 
as a power of $1/Q^2$ which contradicts the
experimental data.

The generalized vector meson dominance model that we use resembles the
aligned jet model in the sense that the meson-nucleon cross section 
does not depend on the mass of the meson, which preserves the 
scaling of shadowing. The cancelation of diagonal terms by
off-diagonal ones brings the additional factor of  $1/m_{h}^2$
which restores the Bjorken scaling. Therefore, these two properties
exhibited by the GVMD model are possible only when the photon 
has both ''hard'' and ''soft'' components.                                    
The ability of the GMVD model to describe the experimental
data stems from the fact the model takes into account basic coherent 
QCD phenomena such as cross section fluctuations, which arise due to 
non-diagonal transitions, and the Color Transparency Phenomenon, which
was taken into account in the particular choice of non-diagonal terms.

Using the exact form of $\Sigma_{nm}$ (\ref{sigmaexpl}) one can present 
with good accuracy the total photoabsorption cross 
section at $Q^2 > 1$ GeV$^2$:
\begin{equation}
\sigma_{T}(s,Q^2>1 GeV^2)=
\frac{e^2}{f^2_{0}}\sigma_{0}\sum_{n}\frac{M^4_{0}}
{(M^2_{n}+Q^2)^2}\Big(\frac{M^2_{n}}{M^2_{n}+Q^2}+0.3\Big) ,
\end{equation}   
which exhibits $1/Q^2$ behavior.

Using the formalism of cross section fluctuations, one can show that 
the second moment of the cross section in the adopted model is given as

\begin{equation}
\langle \sigma^2(s,Q^2) \rangle=\sum_{nmk} 
\frac{e}{f_{n}} \frac{M_{n}^2}{M_{n}^2+Q^2}\Sigma_{nk}
\Sigma_{km}\frac{e}{f_{m}} \frac{M_{m}^2}{M_{m}^2+Q^2} \label{tot}.
\end{equation}
It turns out that $\Sigma_{nk}\Sigma_{km} \propto 1/M^2_{n}$ for 
large $n$, $m$, $k$, which leads to
\begin{equation}
\langle \sigma^2(s,Q^2) \rangle \propto \frac{1}{Q^2}
\end{equation}
for $Q^2 > 1$ GeV$^2$.

Numerical calculations were carried out for 10 vector mesons. The 
dependence of
$\langle \sigma^2(s,Q^2) \rangle /  \langle \sigma(s,Q^2) \rangle$ on 
$Q^2$ is presented in Fig. 2, which shows that for $Q^2 > 5$ GeV$^2$ 
shadowing becomes independent on $Q^2$.
Note that since our aim in this paper is to investigate soft QCD physics,
we ignore QCD evolution due to hard radiation. At comparison with 
the experimental data QCD evolution should be taken into account.

\section{asymptotic behavior of $P_{\gamma}(\sigma,Q^2)$ at small $\sigma$}

Using the explicit form of $\Sigma_{nm}$ given by 
Eq.\ (\ref{sigmaexpl}), one can write the total 
photoabsorption cross section \ (\ref{ttot}):
\begin{equation}  
\sigma_{T}(s,Q^2)=\sum_{n} \frac{e}{f_{n}} 
\frac{M_{n}^2}{M_{n}^2+Q^2}(\sigma_{0}  
\frac{e}{f_{n}} \frac{M_{n}^2}{M_{n}^2+Q^2}+2c_{n} 
\frac{e}{f_{n+1}} \frac{M_{n+1}^2}{M_{n+1}^2+Q^2}) \label{start},
\end{equation}
where
\begin{equation}
c_{n}=-\frac{\sigma_{0}}{2} \frac{M_{n}}{M_{n+1}}(1-0.3 
\frac{M_{0}}{M_{n+1}}) .
\end{equation}

Numerical studies show that if one wants to study  the limit 
of small $\sigma$, one should increase the number of vector mesons 
included in sum \ (\ref{start}) (the dimensionality of  $\Sigma_{nm}$). 
As a result the eigenvalues (possible cross sections) $\sigma_{n}$ 
fill the interval [0,2$\sigma_{0}$] more and more densely. Therefore, 
the minimal eigenvalue (cross section) decreases with increase of the 
number of vector mesons $L$.  
Note that when $L$ is a large number, $\frac{M_{0}^2}{M_{L}^2+Q^2}$ is
a small parameter, hence the expression for $\sigma_{T}(s,Q^2)$ can be
simplified further:
\begin{equation}
\sigma_{T}(s,Q^2)=2e^2 \frac{M_{0}^2}{f_{0}^2} \sum_{L}  
\frac{M_{L}^2}{(M_{L}^2+Q^2)^2} \sigma_{0} (0.15 
\frac{M_{0}^2}{M_{L}^2}+\frac{M_{0}^2}{M_{L}^2+Q^2}) .
\end{equation}

The ultimate goal of this subsection
is to express $\sigma_{T}(s,Q^2)$ as an integral over cross sections. 
We have explained that large masses $M^2_{L}$ correspond to small
cross sections, therefore we use the following  fit for the calculated
minimal cross section (eigenvalue of $\Sigma_{nm}$) for 
large $M^2_{L}$ (large $L$) :
\begin{equation}
\sigma(M^2_{L})=A\sigma_{0} \frac{M_{0}^2}{M^2_{L}} \label{model}.
\end{equation}
 Numerical works with $\Sigma_{nm}$ reveal that $A$ does not 
depend significantly on $N$ for its large enough values, for 
example, for $N=85$ $A=2.12$, for $N$=100 $A=2.06$, for $N=200$
$A=1.88$. In our numerical analysis we use {\it the averaged} value $A=2$.

Since for large masses (large $L$) the relative difference between 
neighboring cross sections $\sigma(M^2_{L})$ is small, it is accurate 
to replace this sum by an integral over cross sections $\sigma$
(keeping in mind that we integrate over small cross sections):
\begin{equation}
\sigma_{T}(s,Q^2)=\int d\sigma \frac{e^2}{f_{0}^2}\sigma_{0}
\frac{1}{(\sigma / \sigma_{0})^2}\frac{AM^4_{0}}{(AM^2_{0}
\sigma_{0} / \sigma+Q^2)^2} \cdot \Big(0.15+\frac{1}{\sigma / 
\sigma_{0}} \frac{AM^2_{0}}{AM^2_{0}\sigma_{0} / \sigma+Q^2} \Big) .
\end{equation}

According to the definition of $P_{\gamma}(\sigma,Q^2)$:
\begin{equation}
\sigma_{T}(s,Q^2)=\int d \sigma P_{\gamma}(\sigma) \sigma .
\end{equation}
Comparing these two expressions, we notice that that at small $\sigma$:
\begin{equation}
P_{\gamma}(\sigma,Q^2)= \frac{e^2}{f_{0}^2}\frac{1}{(\sigma / 
\sigma_{0})^3}\frac{AM^4_{0}}{(AM^2_{0}\sigma_{0} / 
\sigma+Q^2)^2} \cdot \Big(0.15+\frac{1}{\sigma / 
\sigma_{0}} \frac{AM^2_{0}}{AM^2_{0}\sigma_{0} / \sigma+Q^2} \Big) .
\end{equation}
One can see that this formula supports our statement that
$P_{\gamma}(\sigma,Q^2) \propto 1/ \sigma$ at small $\sigma$. 
It shows 
that small cross sections have a sizable probability to 
exist in the photon. This phenomenon 
leads to Color Transparency Phenomena.

To make numerical estimate of the coefficient of proportionality, we 
present the distribution $P_{\gamma}(\sigma,Q^2)$ as
\begin{equation}
P_{\gamma}(\sigma,Q^2)=e^2\frac{1}{\sigma}I_{VMD}(Q^2) \label{limvmd}, 
\end{equation}
where the coefficient $I_{VMD}(Q^2)$ as a function of $Q^2$ is given 
in Table \ (\ref{Table2}). It was calculated for $\sigma /\sigma_{0}=0.1$ .

There is an alternative approach to the calculation
of the asymptotic (limiting) behavior of $P_{\gamma}(\sigma,Q^2)$
based on  QCD, which presents the answer in terms of the photon 
wave function and the cross section of interaction of a small size 
$q\bar{q}$-configuration \cite{Str1}:      
 
\begin{equation}
P(\sigma,Q^2)=\frac{d b^2}{d \sigma(b^2)}
\int_{0}^{1}\sum_{\lambda_{1},
\lambda_{2}}|
\psi(z,b=0)_{\lambda_{1},\lambda_{2}}|^2\frac{dz}{4}N_{C} 
\label{genfor} .
\end{equation}
Here $\psi(z,\vec{b})_{\lambda_{1},\lambda_{2}}$ is a light-cone wave 
function of the photon made of two quarks with momentum fractions $z$ 
and $1-z$ and helicities $\lambda_{1}$ and $\lambda_{2}$ ;
$\sigma(b^2)$ 
is the cross section of the $q\bar{q}$ pair of the transverse size $b$.

In the momentum space the standard expression for the transversely 
polarized  photon wave function is \cite{Str2}, \cite{BrLe}:
\begin{eqnarray}
\psi(z,\kappa_{\perp})_{\lambda_{1},\lambda_{2}}&=&ee_{f}\frac{1}
{Q^2+\frac{\kappa_{\perp}^2+m^2}{z(1-z)}} \, \frac{1}{z(1-z)} 
\Big(-\delta(\uparrow \uparrow)m(\epsilon^{1}+i \epsilon^{2})+
\delta(\downarrow \downarrow)(\epsilon^{1}-i \epsilon^{2}) \nonumber\\
&+&\delta(\uparrow \downarrow)((1-2z)\vec{\epsilon}\cdot 
\vec{\kappa_{\perp}}+i\vec{\epsilon}\times \vec{\kappa_{\perp}}+
\delta(\downarrow \uparrow)((1-2z)\vec{\epsilon}\cdot
\vec{\kappa_{\perp}}
-i\vec{\epsilon}\times \vec{\kappa_{\perp}}) \Big)
\end{eqnarray}

Here $e_{f}$ is the electric charge of the quark with flavor 
$f$ ; $\kappa_{\perp}$ is the transverse momentum of the 
$q\bar{q}$ pair; $\vec{\epsilon}$ is the polarization vector 
of the photon. The arrows stand for different helicities of the quarks. 

Then we perform a Fourier transform into the impact parameter space: 
\begin{equation}
\psi(z,b)_{\lambda_{1},\lambda_{2}}=\frac{1}{(2\pi)^2} 
\int d^2 \kappa_{\perp} \psi(z,\kappa_{\perp})_{\lambda_{1},
\lambda_{2}}e^{-i \vec{\kappa_{\perp}} \cdot \vec{b_{\perp}}} \label{Fourier1}
\end{equation} 
and study the limit $b \rightarrow 0$. The analysis shows that only 
spin-flip components of the wave function give the leading contribution
 at the limit $b \rightarrow 0$ :
\begin{eqnarray}
\psi(z,b \rightarrow 0)_{\lambda_{1},\lambda_{2}}=
-\frac{i}{2\pi}ee_{f}\frac{1}{b} 
\Big(\delta(\uparrow \downarrow)((1-2z)\vec{\epsilon}\cdot 
\vec{n_1}&+&i\vec{\epsilon}\times \vec{n_1} \nonumber\\
+\delta(\downarrow \uparrow)((1-2z)\vec{\epsilon}\cdot  
\vec{n_1}&-&i\vec{\epsilon}\times \vec{n_1}) \Big),
\end{eqnarray}
where $\vec{n_1}$ is a unit vector in the $x$-direction. At 
once one can see that the photon wave function is singular 
at the limit $b \rightarrow 0$. Namely this property 
leads to a singular behavior of $P_{\gamma}(\sigma,Q^2)$ at small $\sigma$.

Upon averaging over two possible polarizations of the 
photon (not written explicitly) and two possible helicities of 
quarks and integrating over $z$ one arrives at the following expression:
\begin{equation}
\int_{0}^{1}\sum_{\lambda_{1},\lambda_{2}}| \psi(z,b 
\rightarrow 0 )_{\lambda_{1},\lambda_{2}}|^2 dz=e^2 e_{f}^2 
\frac{1}{b^2} I^{\prime}_{QCD}(Q^2) \label{wf} ,
\end{equation}
where $I^{\prime}_{QCD}(Q^2)$ is some numerical factor.

The effective cross section of a $q\bar{q}$-- pair interacting with a
nucleon, $\sigma(b^2)$, as a function of the impact parameter $b$, is
given by \cite{Str1}: 
\begin{equation}
\sigma(b^2)=\frac{\pi^2}{3}b^2
xG_{T}(x,\tilde{Q}^2)\alpha_{s}(\tilde{Q}^2) 
\label{cross}.
\end{equation}

For $x=10^{-3}$ the effective energy $\tilde{Q}^2=9/b^2$ \cite{Koepf}.
For small but finite cross sections and, therefore, small $b^2$, the 
combination $xG_{T}(\tilde{Q}^2)\alpha_{s}(\tilde{Q}^2)$ weakly 
depends on $\tilde{Q}^2$. We do not consider this effect in the paper. 

Therefore,  at small $b^2$, one can write:
\begin{equation}
\frac{d \sigma(b^2)}{d b^2}=
\frac{\pi^2}{3} xG_{T}(x,\tilde{Q}^2)\alpha_{s}(\tilde{Q}^2) \label{deriv}.
\end{equation}
Combining equations \ (\ref{genfor}), \ (\ref{wf}) and \ (\ref{deriv})
we obtain at the limit $\sigma \rightarrow 0$:
\begin{equation}
P_{\gamma}(\sigma,Q^2)=e^2\frac{1}{\sigma}I_{QCD}(Q^2) \label{qcdpred},
\end{equation}
where the coefficient $I_{QCD}(Q^2)$ as a function of $Q^2$ is given
in Table \ (\ref{Table2}). 
It was calculated for $\sigma /\sigma_{0}=0.1$ to make a comparison
with the similar prediction of GVMD.

One can see from Table \ (\ref{Table2}) that both GVMD model and QCD give
answers of the same order of magnitude.
Since the wave function of $\rho$-mesons 
is $\frac{1}{\sqrt 2}(|u \bar{u} \rangle -|d \bar{d} \rangle)$, one 
should use $e_{f}^2=1/2$ when summing over quark flavors.

Comparing Eqs.  (\ref{qcdpred}) and  (\ref{limvmd}), one can see that
 both GVMD model and QCD predict the singular behavior of $P(\sigma)$ 
at $\sigma \rightarrow 0$: $P_{\gamma}(\sigma) \propto 1/\sigma$. Thus
we conclude that the GVMD model exhibits the phenomenon of Color 
Transparency known from QCD, which means that in the photon, with a 
noticeable probability, there are hadronic configurations weakly 
interacting with the target.

\section{ Hard diffractive electroproduction of photons}

Another application of GVMD which is important for the interpretation 
of hard diffractive phenomena at HERA can be found in 
the diffractive photoproduction in deep inelastic scattering
initiated by highly virtual photon:
$\gamma^{\ast}(Q^2)+ p \rightarrow \gamma^{\ast}(Q_{f}^2) +p$. 

Within GVDM  we found that the ratio of the imaginary 
parts of the amplitudes for real photon 
and virtual photon production is:
\begin{eqnarray}
\frac{{\rm Im}A(\gamma^{\ast}(Q^2)+p\rightarrow \gamma(Q^2=0)+p)}{{\rm Im}A(\gamma^{\ast}(Q^2)+p\rightarrow \gamma^{\ast}(Q^2))+p)}&=& \nonumber\\
\Big(\sum_{nm} \frac{e}{f_{n}}\frac{M_{n}^2}{M_{n}^2+Q^2}\Sigma_{nm}\frac{e}{f_{m}}\Big) &\Big/& \nonumber\\
 \Big(\sum_{nm} \frac{e}{f_{n}}\frac{M_{n}^2}{M_{n}^2+Q^2}\Sigma_{nm}\frac{e}{f_{m}} \frac{M_{m}^2}{M_{m}^2+Q^2}\Big) .
\end{eqnarray}
This ratio is given in Fig. 3. One can see that it grows with $Q^2$. 
Although PQCD predicts that this ratio is approximately
$Q^2$-independent 
at small $x$ and estimates it to
be $\approx$2 \cite{freund}, for a wide range of $Q^2$ both 
answers are rather close.

\section{Conclusions}

We have shown that, using the generalized vector meson dominance
model, one can reconstruct the distribution function 
$P_{\gamma}(\sigma,Q^2)$, which gives the probability for the 
photon to interact with a target with the cross section $\sigma$. 
The very existence of such a distribution supports the idea that 
the photon consists of $q\bar{q}$ configurations  of a different 
geometrical size even within nonperturbative QCD regime,
therefore they interact with the target with different cross sections.

We have studied the limiting behavior of $P_{\gamma}(\sigma,Q^2)$ at 
$\sigma \rightarrow 0$ and shown that $P(\sigma,Q^2) \propto 1/\sigma$.
That is a very remarkable property which means that there is a
sizable probability of finding a small configuration in the photon. 
Moreover, the relative probability to find such a small configuration 
in the photon increases when the photon virtuality $Q^2$ grows.

We conclude that the GVMD model describes 
well Bjorken scaling and nuclear shadowing because it takes into 
account basic coherent QCD phenomena: non-diagonal transitions result 
in cross section fluctuation of the virtual photon; the specific type 
of the non-diagonal terms, chosen in line with the parton model, leads
to a significant probability of spatially small configurations in the 
photon wave function and the
Color Transparency Phenomenon.

As in the parton model, we found in GVMD that although the 
contribution
of soft QCD in the photon wave function decreases with increase of
$Q^2$, it gives a comparable contribution into the total cross section and the
dominant contribution into moments of $P(\sigma,Q^2)$.

Since the model gives  approximate Bjorken scaling of 
nuclear shadowing, it means that the photon has ''hard'' 
and ''soft'' components, which was suggested by the aligned jet model 
and affirmed later by QCD. This reveals the duality between the 
parton language and GVMD description of the photon hadronic structure.   

Also within the GVMD model we give predictions for the ratio 
of the imaginary parts of the amplitudes
of the process $\gamma^{\ast}(Q^2)+ p \rightarrow \gamma^{\ast}
(Q_{f}^2) +p$, which turns out to be consistent with PQCD over 
a wide range of $Q^2$.

\section{acknowledgments}
We thank G. Shaw for very helpful discussions. 

We are thankful to D. Schildknecht for pointing out reference \cite{Schildknecht}.

 This work was 
supported by the U.S. Department of Energy grant number DE-FG02-93ER-40771 and US-Israeli Bi-National Science Foundation Grant number 9200126.

\bibliographystyle{unsrt}

\begin{table}
\begin{tabular}{|c|c|c|c|c|c|}
Moments & $Q^2$=0 & $Q^2$=0.5 GeV$^2$ & $Q^2$=1 GeV$^2$  & $Q^2$=2
GeV$^2$ &  $Q^2$=5 GeV$^2$ \\
\hline 
$\langle \sigma \rangle$  & 1.1421 & 0.4529 & 0.2755 & 0.1519 & 0.0603  \\
$\langle \sigma^2 \rangle$ & 0.8589 & 0.2749 & 0.1452 & 0.0679 & 0.0227 \\
$\langle \sigma^3 \rangle$ & 0.7547 & 0.2135 & 0.1043 & 0.0449 & 0.0142 \\
$\langle \sigma^4 \rangle$ & 0.7180 &  0.1878 & 0.0883 & 0.0374 & 0.0126 \\
$\langle \sigma^5 \rangle$ & 0.7203 & 0.1784 & 0.0829 & 0.0362 & 0.0138 \\
\end{tabular}
\caption{Moments of the distribution $P(\sigma)$ as a function of $Q^2$}
\label{Table1}
\end{table}

\begin{table}
\begin{tabular}{|c|c|c|c|}
 & $Q^2$=0 & $I_{VMD}(Q^2)$ & $I_{QCD}(Q^2)$ \\
\hline
 & 0 & 1.7$\times 10^{-2}$ & 2.8$\times 10^{-2}$ \\
 & 1 & 1.4$\times 10^{-2}$ & 2.5$\times 10^{-2}$ \\
 & 2 & 1.1$\times 10^{-2}$& 2.3$\times 10^{-2}$\\
 & 5 & 0.6$\times 10^{-2}$ & 1.9$\times 10^{-2}$\\
\end{tabular}
\caption{Coefficients $I_{VMD}(Q^2)$ and $I_{QCD}(Q^2)$ as a function of $Q^2$}
\label{Table2}
\end{table}

\begin{figure}
\centering
\mbox{\epsfig{file=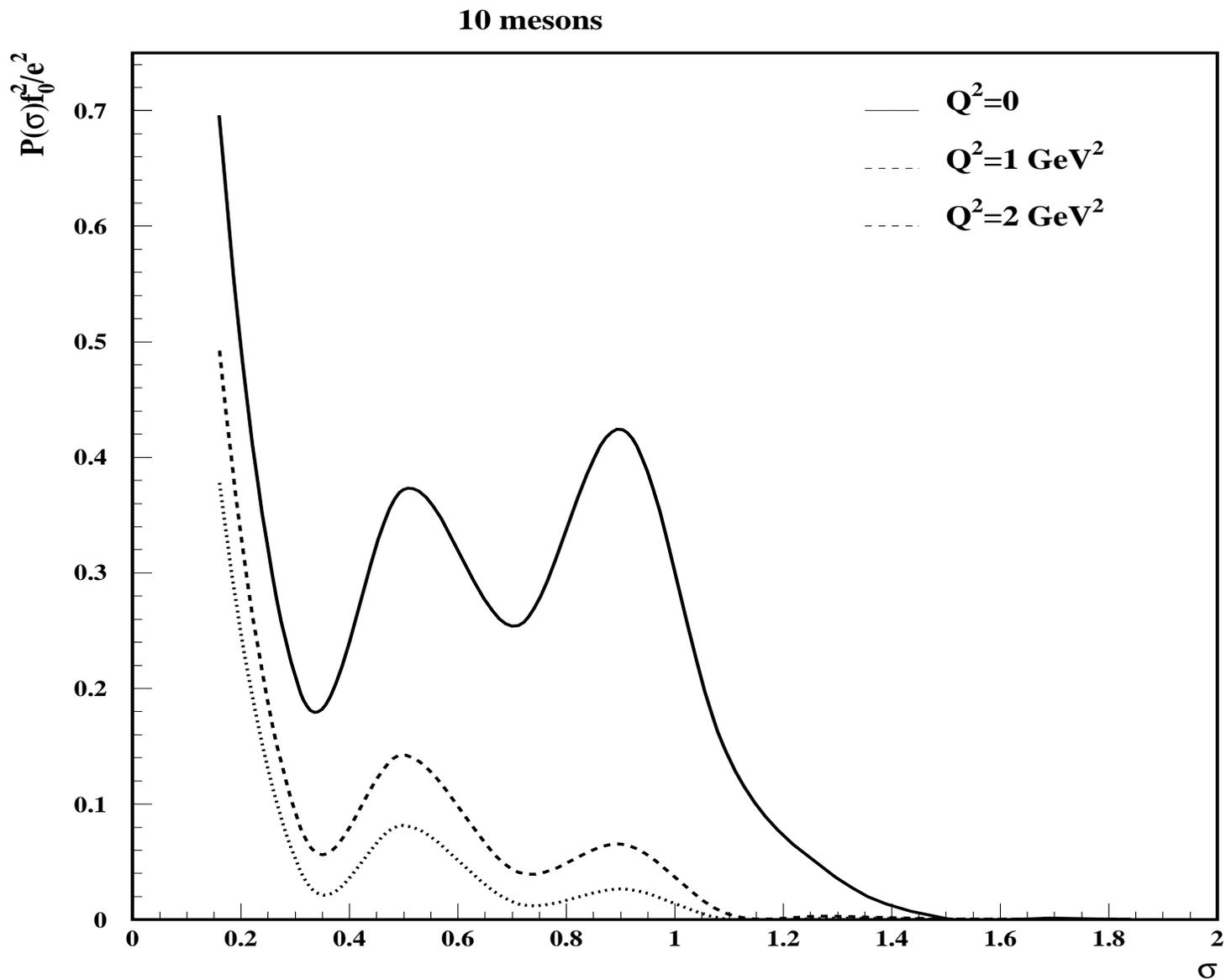,height=20cm,width=17cm}}
\caption{The distribution $P_{\gamma}(\sigma,Q^2)$ as a function of $Q^2$ (we plot $f_{0}^2/ e^{2} \, P(\sigma,Q^2$))}
\label{Fig. 1}
\end{figure}

\begin{figure}
\centering
\mbox{\epsfig{file=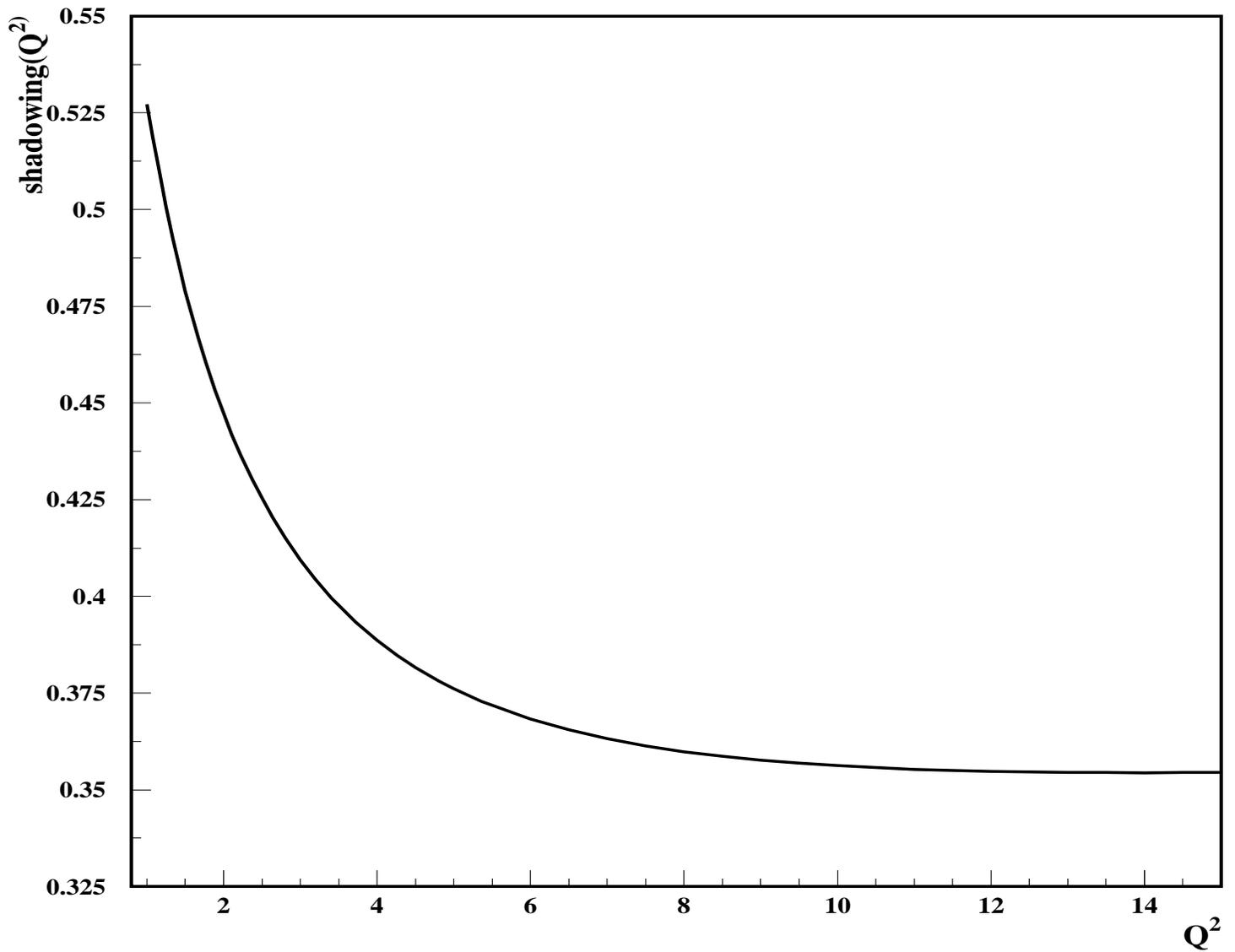,height=20cm,width=17cm}}
\caption{The shadowing factor $shadowing(Q^2)=\langle \sigma^2 \rangle / (\langle \sigma \rangle \sigma_{0})$ as a function of $Q^2$. $\sigma_{0}$  is the total cross section of interaction of vector mesons with the target. }
\label{Fig. 2}
\end{figure}

\begin{figure}
\centering
\mbox{\epsfig{file=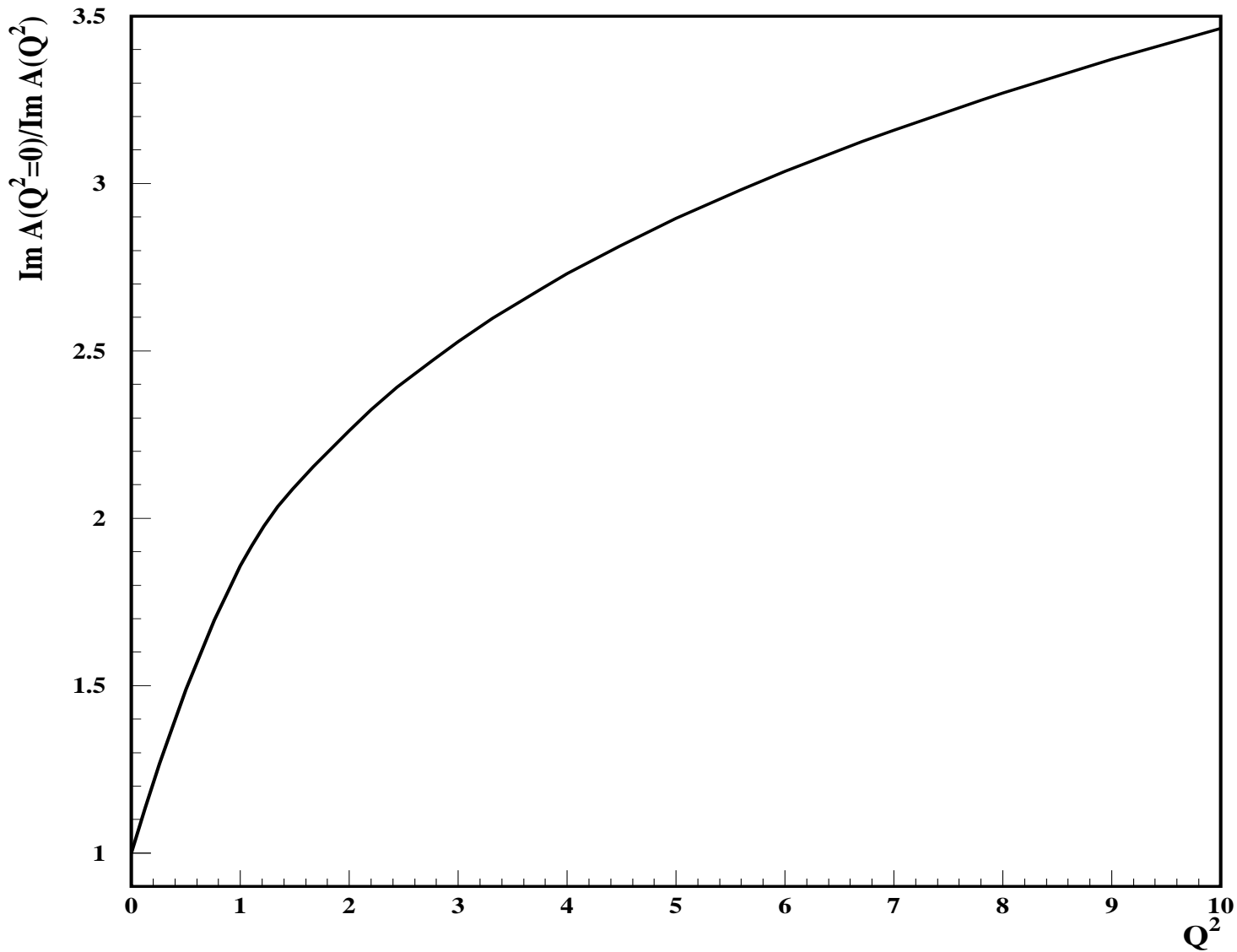,height=20cm,width=17cm}}
\caption{The fraction of the imaginary parts of the amplitudes of real and virtual photon production as a function of $Q^2$. }
\label{Fig. 3}
\end{figure}
\end{document}